# Giant Gating Tunability of Optical Refractive Index in Transition Metal Dichalcogenide Monolayers

Yiling Yu[1,2], Yifei Yu[1], Lujun Huang[1], Haowei Peng[3], Liwei Xiong[1,4], and Linyou Cao[1,2]*

[1]Department of Materials Science and Engineering, North Carolina State University, Raleigh NC 27695; [2]Department of Physics, North Carolina State University, Raleigh NC 27695; [3]Department of Chemistry, Temple University, Philadelphia PA 19405; [4]Hubei Key Laboratory of Plasma Chemistry and Advanced Materials, Wuhan Institute of Technology, Wuhan, PR China, 430205.

**Abstract**

We report that the refractive index of transition metal dichacolgenide (TMDC) monolayers, such as $MoS_2$, $WS_2$, and $WSe_2$, can be substantially tuned by > 60% in the imaginary part and > 20% in the real part around exciton resonances using CMOS-compatible electrical gating. This giant tunablility is rooted in the dominance of excitonic effects in the refractive index of the monolayers and the strong susceptibility of the excitons to the influence of injected charge carriers. The tunability mainly results from the effects of injected charge carriers to broaden the spectral width of excitonic interband transitions and to facilitate the interconversion of neutral and charged excitons. The other effects of the injected charge carriers, such as renormalizing bandgap and changing exciton binding energy, only play negligible roles. We also demonstrate that the atomically thin monolayers, when combined with photonic structures, can enable the efficiencies of optical absorption (reflection) tuned from 40% (60%) to 80% (20%) due to the giant tunability of refractive index. This work may pave the way towards the development of field-effect photonics in which the optical functionality can be controlled with CMOS circuits.

* To whom correspondence should be addressed.

Email: lcao2@ncsu.edu

Electrically gating photons with complementary metal–oxide–semiconductor (CMOS) circuits has the potential to revolutionize the photonic industry. It would enable the development of unprecedented dynamic photonics with spatial and temporal resolutions comparable to that of state-of-art CMOS circuits. Of particular interest is the control of photons in the visible range as it would revolutionize the fields of imaging, cloaking, superlensing, and virtual reality. However, the electrical gating of photons is very challenging because of the difficulty in gating refractive index. Photons cannot be directly manipulated by electric fields because of charge neutrality and may only be controlled by virtue of light-matter interactions, including reflection, transmission, absorptions, and scattering. As refractive index stands as the most fundamental measure for light-matter interactions, to gate photons is in essence to gate refractive index. Numerous approaches have been reported to electrically tune refractive index, including plasma dispersion effect of free charges[1-6], electro-absorption effects (quantum confined stark effect[7] and Keldysh effect[5, 8]), and non-linear electro-optic effects such as Kerr[9] or Pockels effects[10]. But none of these approaches may provide satisfactory tuning efficiency, speed, spectral range, spatial resolution, and compatibility with CMOS circuits at the same time. For instance, the optical absorption of injected charge carriers may give rise to changes in the refractive index, but the density of the charge carriers that can be injected using the conventional CMOS-gating is limited. This fundamentally limits the tunability in refractive index at the scale of 0.1-1% [1-2, 5, 11] or the tuning spectrum at mid-IR or GHz, THz frequencies[3-4]. While ionic gating has been recently reported able to inject higher densities of charges, which gives rise to larger tunability in visible frequencies,[12-14] its nonlocal nature, inferior chemical stability, and intrinsically slow switching limits the operation speed, footprint, and compatibility with CMOS circuits. Additionally, electro-absorption effects, including Keldysh and quantum confined stark effects,[7-8, 15] have been reported able to induce change in optical absorption with the presence of electric fields in the vertical direction,[7-8, 15] but most of these works are in the infrared or even lower energy ranges, the tuning efficiency is often small (no more than few percent), or the structure is too bulky for integration with CMOS circuits.

Here we demonstrate that the refractive index of transition metal dichalcogenide (TMDC) monolayers, including $MoS_2$, $WS_2$, and $WSe_2$, can be substantially tuned by > 60% in the imaginary part and > 20% in the real part around excitonic resonance using CMOS-compatible electrical gating. This large tunability is achieved by leveraging on the strong excitonic effects of the monolayers [16] and the high susceptibility of the excitons to the influence of free charge carriers. We further elucidate that the giant tunability in refractive index is mainly due to the effects of the injected charge carriers in broadening the spectral width of excitonic transitions and facilitating the interconversion of neutral and charged excitons. In contrast, the other effects of the injected charge carriers, such as renormalizing the bandgap and changing the exciton binding energy, play only negligible roles. It is worthwhile to point out that although many previous studies have reported electrical tunability in the light absorption and emission of TMDC monolayers,[17-22] this work is the first ever quantitatively demonstrating the tunability in refractive index and unambiguously elucidating the underlying physics. Additionally, we use a simple design to illustrate that, when combined with optical resonant structures, the tunability of refractive index in the atomically thin monolayers may enable substantial change in light reflection and absorption.

We start with examining the spectral reflectance (the intensity ratio of the light reflected from the monolayer and the light reflected from a mirror) of TMDC monolayers under electrical gating. Figure 1a inset schematically illustrates the measurement configuration. The monolayers (obtained from 2Dlayer) are grown on degenerately doped Si substrates with thermally grown silicon oxide （$SiO_2$/Si） using chemical vapor deposition (CVD) processes. The source and drain electrodes (5 nm Ti/ 50 nm Au) are fabricated on top of the monolayer using standard e-beam lithography and metallization procedures. We have confirmed that the electrodes form Ohmic contact with the monolayers (Fig S5b inset). In experiments, the Si substrate is used as the gate and identical potentials are applied to the source and drain electrodes. Fig. 1a shows the spectral reflection collected from monolayer $WS_2$ under different gating voltages. The reflection of the *A* exciton (~ 1.95eV) shows substantial variation with the gating voltage, but the *B* (~ 2.35eV) and *C* excitons ( ~ 2.70 eV, Fig. S1) show much less or even negligible variation. The appearance of the gated variation only at the excitonic peaks suggests that this is not caused by the plasma dispersion effect of free charge carriers, which would give rise to changes over a broad spectral range. Similar gating tunability in reflectance can also be observed at monolayer $MoS_2$ and $WSe_2$ (Fig. S2). The larger tunability in magnitude for $WS_2$ likely stems from its larger excitonic absorption compare to $MoS_2$ and $WSe_2$ [23].

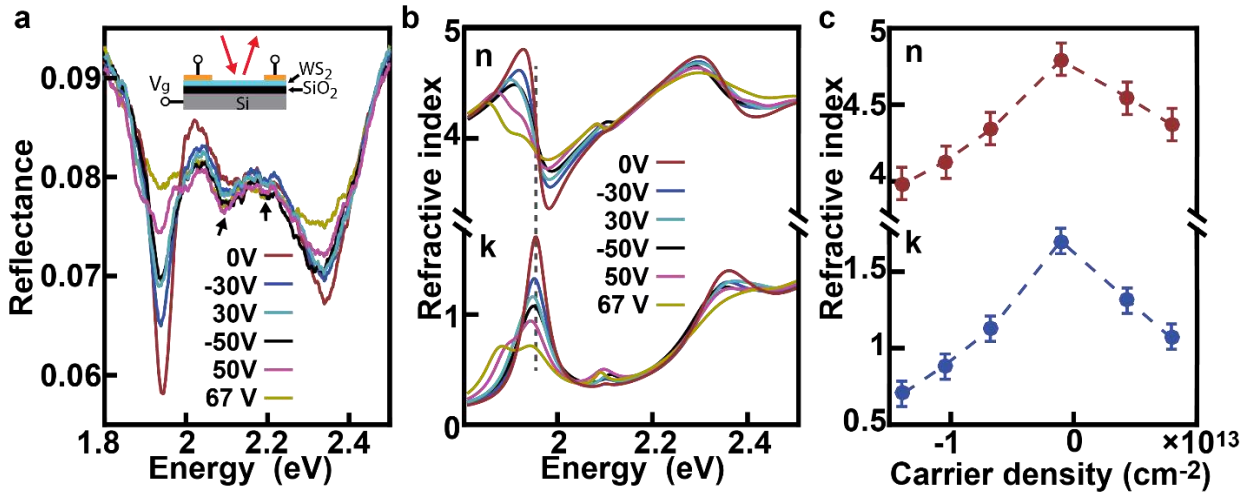

Fig.1 Gated reflection and refractive index of monolayer $WS_2$. (a) Reflection spectra of monolayer $WS_2$ at different gating voltages. The inset is schematic illustration for the measurement configuration. The two arrows point out the two excited states of the A exciton, 2S and 3S. (b) Fitted real part *n* and imaginary part *k* of refractive index with different gating voltages. (c) The peak value of the real part (at around 1.92 eV) and the imaginary part (at around 1.95 eV) as a function of carrier densities.

We can obtain refractive index from the measured spectral reflection using a Kramers-Kronig constrained analysis. Basically, we fit the complex dielectric function $\varepsilon = \varepsilon_1 + i\varepsilon_2$ of the monolayer as a sum of contribution from multiple Lorentz oscillators[24-25]:

$$\varepsilon(E) = \varepsilon_\infty + \sum_j \frac{f_j}{E_j^2 - E^2 - iE\gamma_j} \quad \ldots (1)$$

where $\varepsilon_\infty$ is the high frequency dielectric constant, $E_j$ and $\gamma_j$ are the resonant energy and damping factor of the $j$th oscillator, respectively. $f_j$ is a phenomenal parameter involving contribution from transition matrix element, density of states, and excitonic effects. Additionally, we fit the reflection of the monolayer on top of $SiO_2$/Si substrates using a model for the reflection of multilayer structures as[26]

$$R = |r|^2 = \left| \frac{r_1 + r_1 e^{-i2k_1 d_1} + r_1 r_2 r_3 e^{-i2k_2 d_2} + r_3 e^{-i2(k_1 d_1 + k_2 d_2)}}{1 + r_2 r_1 e^{-i2k_1 d_1} + r_2 r_3 e^{-i2k_2 d_2} + r_1 r_3 e^{-i2(k_1 d_1 + k_2 d_2)}} \right|^2 \quad \ldots (2)$$

where $\rho_1 = (n_1 - n_0)/(n_1 + n_0)$, $\rho_2 = (n_2 - n_1)/(n_1 + n_2)$, and $\rho_3 = (n_3 - n_2)/(n_2 + n_3)$. $n_0$, $n_1$, $n_2$, and $n_3$ are the refractive index of air, the monolayer, $SiO_2$, and Si, respectively. $k_1$ and $k_2$ are the wavenumber in the monolayer and $SiO_2$ as $k_1 = n_1 2\pi/\lambda$ and $k_2 = n_2 2\pi/\lambda$. $d_1$ and $d_2$ are the thickness of the monolayer and the $SiO_2$. The refractive index of the monolayer $n_2$ can be correlated to the fitted dielectric function $\varepsilon$ as $n_2^2 = \varepsilon$. In order to get accurate determination of the dielectric function $\varepsilon$, the Kramers-Kronig constrained analysis requires information in the full spectral range, but the measured spectral reflection only cover the range of 1.8-2.5 eV. To address this issue, we ignore the contribution from the oscillators in lower energy ranges as it is expected to be weak for the refractive index in the visible range. However, the contribution from the oscillators at higher energy ranges has to be considered. We assume that the dielectric function of the monolayer at the higher energy ranges is similar to that of bulk counterparts, and use the dielectric function of the bulk counterparts, which is available in reference[27], to correct the oscillators of the monolayers in the higher energy range. This strategy has been previously demonstrated able to give rise to reasonably accurate dielectric function of TMD monolayers from spectral reflectance[23]. We have also confirmed that the refractive index obtained using this strategy is indeed consistent with what measured from spectral ellipometry, the standard characterization techniques for dielectric functions (Fig. S8). More detailed description about the fitting process can be found at the Supporting Information. Three major oscillators are involved in the measured spectral reflection, including the neutral $A$ exciton ($A_0$), the charged $A$ trion ($A_-$ or $A_+$), and the $B$ exciton. The resonant energy $E_j$ and damping factor $\gamma_j$ for each of the oscillators can be found out from the measured spectral reflection (Fig. 1a). There are another two small oscillators corresponding to the excited states (2S and 3S) of the $A$ exciton as indicated by the black arrows in Fig. 1a,[28] but they can be ignored due to trivial contribution.

Fig. 1b shows the refractive index obtained from the analysis. The fitted reflection spectra match the experimental results very well (Fig. S3). Additionally, the sum rule holds for the fitted reflective index under all the different gating voltages [29] as the integration of the fitted absorption coefficient $\alpha(\omega)$ over the full spectrum range (up to 6 eV) always gives rise to similar values regardless the gating voltage (Fig. S4). All these further support the validity of the fitting method. The refractive index at the frequencies around the $A$ exciton shows substantial tunability, the real part tuned from 4.80 to 3.97 and the imaginary part from 1.7 to 0.7 when the gating voltage $V_g$ is changed from 0 to 67 V. This is two orders of magnitude higher than what previously reported for tuning the refractive index in the visible range by electrical gating.[11] Note that the maximal tunability of the real and imaginary parts appears at different frequencies, 1.95 eV for the imaginary part and around 1.92 eV for the real part. This is rooted in the conjugation nature of the real and imaginary parts of dielectric functions. To further illustrate the tunability, we plot the real part of the refractive index at 1.92 eV and the imaginary part at 1.95 eV as a function of the density of charge carriers in Fig.1c. The charge density is estimated using the capacitor model $Q = C_{ox}(V_g-V_{th})$, in which $C_{ox}$ is the oxide capacitance, $V_g$ is the gate voltage, and $V_{th}$ is the threshold voltage for charge neutrality in the monolayer that we find to be around -7V from PL[18] and I-V measurements (Fig. S5). Both the real and imaginary parts of the refractive index show a maximum at the point of charge neutrality and decrease with the density of charge carriers (either electrons or holes) increasing.

The observed tunability in refractive index can be mainly correlated to the gated variation in the absorption of the neutral $A$ exciton ($A_0$). This is evidenced by the imaginary part of the dielectric function $\varepsilon_2$ (Fig. S6), which is known proportional to optical absorption and may uniquely determine the real part of the dielectric function via the Kramers-Kronig relationship.[25] The fitting result for $\varepsilon_2$ indicates that only the absorption of the neutral $A$ exciton ($A_0$) shows substantial variation with electrical gating. For the convenience of discussion, we only focus on the on-resonance absorption of $A_0$ that is proportional to $\varepsilon_2 = f_{A0}/(E_{A0} \cdot \gamma_{A0})$. The resonant frequency $E_{A0}$ does not change much with the gating voltage (Fig. 2a). Therefore, the observed tunability in refractive index is essentially dictated by the variation of $f_{A0}$ and $\gamma_{A0}$ under electrical gating. To better illustrate this notion, we plot $\gamma_{A0}$ (obtained from Fig. 1a) and $f_{A0}$ (obtained from the Kramers-Kronig constrained analysis with eq. (1)) as a function of charge carrier densities in Fig. 2b-2c. The $f$ and $\gamma$ of the charged $A$ exciton ($A_{+/-}$) are also plotted in Fig.2 as reference.

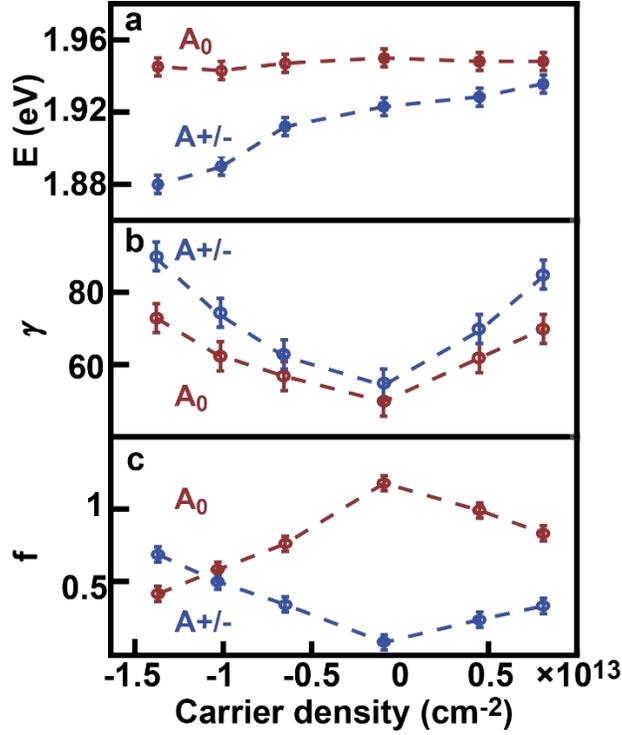

Fig.2 Gating tunability of physical parameters in monolayer $WS_2$. (a) Optical bandgap $E$ (b) damping factor $\gamma$, and (c) $f$ that representing the oscillation strength of the neutral ($A_0$) and charged ($A_{+/-}$) excitons. The error bar given in the figure represents the error in fitting the measured spectral reflectance.

We may better understand the physics underlying the gated variation of $f_{A0}$ and $\gamma_{A0}$ through analysis for the effect of the injected charge carriers. Generally, the injected charge carriers may affect the optical properties of low-dimensional semiconductor materials through three major physical mechanisms: Pauli blocking, Coulomb scattering, and dielectric screening. The Pauli blocking causes phase space filling up to Fermi level (Table. S1), which will lead to blue shift of the optical gap energy. The Coulomb scattering, through which excitons interact with the injected charge carriers, may broaden the spectral width of excitonic absorption by enhancing the dephasing rate and also facilitate the formation of charged excitons. The screening of Coulomb interactions may lead to bandgap renormalization and change in the exciton binding energy.[28, 30-31] The bandgap renormalization and the change in exciton binding energy can be estimated from Raman measurements. The intensity ratio of the two characteristics Raman peaks of monolayer WS$_2$, E$_{2g}$/2LA(M) and A$_{1g}$, decreases with the charge carrier density increasing (Fig. 3a). This can be correlated to the bandgap renormalization as the intensity of the E$_{2g}$/2LA(M) peak is related with the band structure due to the involvement of double resonances.[32] We can estimate the amplitude of the bandgap renormalization by comparing the Raman spectra measured at different gating voltages to those collected under different temperatures. This is because the Raman intensity ratio decreases with the temperature increasing, and the temperature dependence is similar to the dependence of the intensity ratio on the gating voltage (Fig. 3a). Briefly, we identify the temperature under which the Raman intensity ratio is comparable to what observed at specific gating voltages, and then estimate the bandgap renormalization based on a well-established temperature-bandgap correlation (Fig.3b).[33] For simplicity, the effect of temperatures on the exciton binding energy is ignored, which is reasonable given the relative small temperature change in our experiments (300K to 450K). The estimated bandgap renormalization $\Delta E_g$ (using the bandgap at the gating voltage of 0V as reference) is plotted as a function of the density of charge carriers in Fig. 3c. With the information of bandgap renormalization $\Delta E_g$, we can derive the change in exciton binding energy $\Delta E_{ex}$ from the change in the optical bandgap $\Delta E_{opt}$ as $\Delta E_{opt} = \Delta E_g - \Delta E_{ex} + \Delta E_F$. E$_F$ here is the fermi energy that indicates the phase space filling effect. For the neutral $A$ exciton, $\Delta E_{ex}$ is smaller than 30mev in our gated voltage range ($\Delta E_F$ and $\Delta E_g$ can get from Table S1, Fig. 2a). By the same token, we may also estimate $\Delta E_{ex}$ of the charged $A$ exciton (see Table S1).

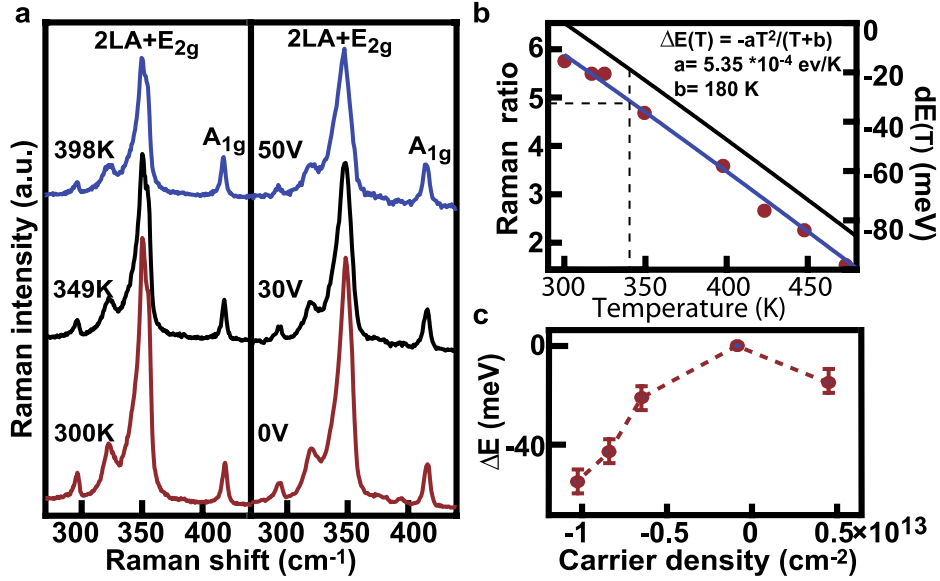

Fig.3 Bandgap renormalization and change in exciton binding energy in monolayer $WS_2$. (a) Comparison of Raman spectra of monolayer $WS_2$ at different temperatures (left) and different gating voltages (right). The intensity is normalized to that of the $A_{1g}$ peak. (b) The measured (red dots) and fitted (blue line) Raman intensity ratio as a function of the temperature (red dots). The black curve is the calculated bandgap shift of the monolayer as a function of temperature, in which the bandgap at 300 K is used as the reference. The dashed lines indicate one Raman intensity ratio measured at the monolayer under electrically gating. (c) The estimated bandgap renormalization of monolayer $WS_2$ as a function of the density of charge carriers.

It is obvious that the Coulomb scattering is the major mechanism for the gated variation in the damping factor $\gamma$. We can find out the mechanism governing the gated variation of $f$ through fitting the dielectric function with a model of Wannier excitons in fractional dimension space. The fractional dimensional space model has previously been established to describe the optical properties of quantum wells. [34-39] It treats highly anisotropic excitons in low-dimensional structures to be isotropic in a fractional-dimensional space, and defines an effective dimensionality $d$ based on the excitonic binding energies in bulk materials $R$ and the low-dimensional structure $E_{ex}$ as $d = 1+2(R/E_{ex})^{0.5}$. One major advantage of the fractional dimension space model lies in its capability to quantitatively evaluate the contribution of excitonic effect, which is realized by introducing the effective dimensionality to the calculation of joint density of states. According to the fractional dimension space model, the complex dielectric function $\varepsilon_j$ at an arbitrary energy $E$ contributed by one excitonic transition can be written as $\varepsilon_j = S_j \bullet G_j$, where $S_j$ is a parameter representing the transition probability and $G_j$ is an effective joint density of state in which the excitonic effect is represented by the effective dimensionality $d$. [34-39] More specifically,

$$S_j = F|p_{vc}|^2$$
$$G_j = \frac{R^{d/2-1}}{(E+i\gamma_j)^2}\left[g_d(\xi(E+i\gamma_j)) + g_d(\xi(-E-i\gamma_j)) - g_d(\xi(0))\right] \quad \ldots (3)$$
$$g_d(\xi) = \frac{2\pi\Gamma((d-1)/2+\xi)}{\Gamma((d-1)/2)^2\Gamma(1-(d-1)/2+\xi)\xi^{d-2}}\left[\cot\pi((d-1)/2-\xi) - \cot\pi(d-1)\right]$$

where $F$ is a constant prefactor, $p_{vc}$ is the matrix element for the interband transition, $\mu$ is the reduced effective mass of electrons, $e$ and $m_0$ are the charge and mass of free electrons, $\varepsilon_0$ and $\hbar$ are the vacuum permittivity and the Planck's constant, $\Gamma()$ is the gamma function, $\xi()$ is a function related with the electronic bandgap $E_g$ and $R$ as $\xi(z) = [R/(E_g - z)]^2$. The exciton binding energy in monolayer and bulk $WS_2$ is set to be 0.7 eV [40-41] and 0.055 eV [27], respectively. The other parameters, including the effective dimensionality $d$, electronic bandgap $E_g$, damping factor $\gamma$ can be obtained from either the known binding energy, the experimental measurement, or the preceding discussion. We use the eq. (3) to fit the dielectric function. Basically, we first calculate $G_j$ using the known parameters and then fit the value of $S_j$ to match the measured reflection spectra (Fig. S7).

The fitting result indicates that the gated variation in $f_{A0}$ mainly results from the interconversion of neutral and charged $A$ excitons. Fig. 4a shows the fitted $S_j$ for the neutral ($A_0$) and charged ($A_{+/-}$) excitons at different carrier densities. We can get useful physical insight by comparing $S_j$ to $f_j$ as shown in Fig. 4a. For the convenience of comparison, all the results are normalized to the sum of the values for both neutral and changed excitons ($S_{A0} + S_{A+/-}$ and $f_{A0} + f_{A+/-}$) at each gating voltages respectively. The $S_j$ and $f_j$ show very similar dependence on the charge carrier density. This indicates that the enhancement by excitonic effects to the oscillation strength, which is involved in $f_j$ but not in $S_j$, does not change much with the gating voltage. The variation in $f_j$ under electrical gating mainly results from change in the transition matrix element $p_{vc}$, instead of change in the excitonic effect. Additionally, the sum of the $S_j$ for the neutral and charged excitons together remains to be reasonably constant regardless the gating voltage. This suggests that the interband transitions of the neutral and charged excitons are competing processes that involve the same ground state. The change of the $S_j$ for the neutral and charged excitons with gating voltages results from the interconversion between the neutral and charged excitons, which is indicated by the redistribution of the transition matrix element $p_{vc}$ among the neutral and charged excitons.

The correlation of gated variation in $f_j$ to the interconversion of neutral and charged excitons may be further supported by the thermal equilibrium distribution of the excitons. We use the negatively charged exciton as an example to illustrate this notion. The interconversion of the neutral and negatively charged exciton can be written as $A_0 + e \leftrightarrow A^-$. Following what has been previously studied,[42-43] we assume that, in the temperature and photoexcitaton range of our experiments, the monolayer consisting of the neural and charged excitons behaves as a two-level system in equilibrium. The thermal equilibrium distribution of the neutral and charged excitons is dictated by the chemical potential of the injected electrons $\xi$ and the binding energy of the charged exciton $E_b^{A-}$. The chemical potential of a two-dimensional ideal fermion gas can be written as $\xi = k_B T ln(e^{E_F/k_B T} - 1)$, where $k_B$ and $T$ are the Boltzmann constant and temperature, respectively. $E_F$ is the Fermi energy of the injected electrons at the conduction band and can be calculated with $E_F = \pi \hbar^2 n_e/(2m_e)$, where $n_e$ and $m_e$ are the density and effective mass of the injected electrons, and the constant of 2 stands for valley degeneracy of the monolayer.[44] The ratio between the densities of the neutral ($n_{A0}$) and charged ($n_{A-}$) excitons under thermal equilibrium is a function of the binding energy of charged exciton $\frac{n_{A0}}{n_{A-}} = 4e^{(-E_b^{A-}-\xi)/k_B T}$, where the factor 4 is the degeneracy ratio of the neutral and charged excitons.[42] Therefore, the fractions of the neutral and charge excitons are defined as $r_{A0} = n_{A0}/(n_{A0} + n_{A-})$ and $r_{A-} = n_{A-}/(n_{A0} + n_{A-})$. We can estimate the binding energy $E_b^{A-}$ from the measured optical bandgaps of the neural ($E_{A0}$) and charge ($E_{A-}$) excitons and the binding energy as $E_b^{A-} = E_{A0} - E_{A-} - E_F$. Similar analysis can be performed for the positively charge $A$ exciton ($A_+$). With all the information, we can calculate the fractions $r_{A0}$ and $r_{A-}$ as a function of the density of injected charge carriers (see Table S1 in the Supporting Information for the parameters used in the calculation). The calculation results are plotted in Fig. 4a. It shows reasonable consistence with the $f_j$, further indicating the gated variation of $f_j$ can be mainly ascribed to the interconversion of the neutral and charged excitons.

The fitting result confirms that the damping factor $\gamma$ is important for the observed tunability in refractive index. It also indicates that the change in excitonic binding energy $\Delta E_{ex}$ and the bandgap renormalization $\Delta E_g$ play negligible roles. These variables are all included in the parameter $G_j$ but not $S_j$. We examine $G_j$ as a function of each of the variable. The change in the damping factor $\gamma$ dominates the variation in $G_j$, while the change in binding energy $\Delta E_{ex}$ (represented by change in the effective dimensionality $d$) and the bandgap renormalization $\Delta E_g$ (represented by change in the electronic bandgap $E_g$) may only have minor effects (Fig. 4c). Intuitively, this result is understandable as the $\Delta E_{ex}$ and $\Delta E_g$ are more than one order of magnitude smaller than the exciton binding energy and bandgap, respectively.

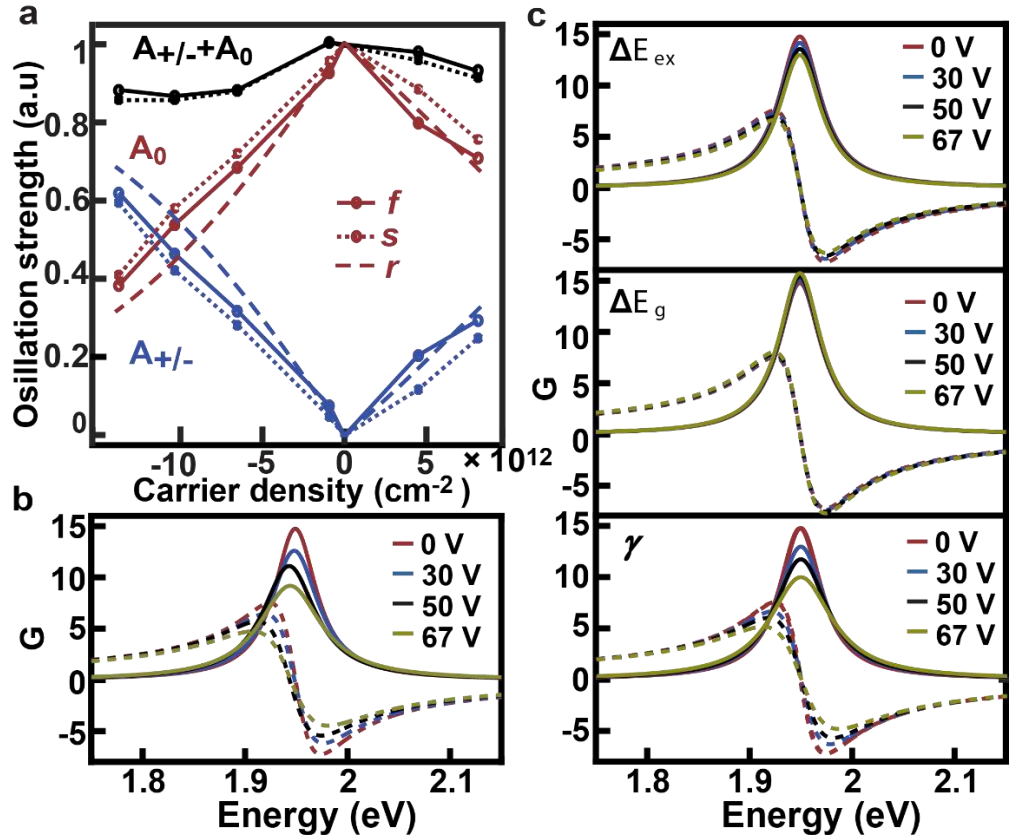

Fig.4 Dominating mechanism for the tunable refractive index. (a) Fitting results for *S* and *f* of the neutral and charged *A* excitons as function of charge carrier densities. The black solid and dashed lines indicate the sum of the values for both neutral and charged excitons. The calculated fractions of the neutral and charged excitons *r* are also given. (b) Calculated *G* for the neutral *A* exciton with different gating voltages. (c) Calculated *G* of the neutral *A* exciton as a function of (upper) change in exciton binding energy, (middle) bandgap renormalization, (c) change in the damping factor *γ*. The spectra in the upper and middle panels are artificially shifted to align the peak position for the convenience of comparison.

As the intrinsic optical response of 2D TMDC materials is weak due to the atomically thin dimension, the development of field-effect photonic devices for practical applications would require substantial enhancement in the optical response, and this can be achieved by leveraging on the power of optical resonance. [45] To illustrate this notion, we have designed a GaN based grating structure on $Al_2O_3$ with monolayer $WS_2$ covered on the top and a silver mirror at the bottom as illustrated in Fig. 5a. The GaN layer is heavily doped serve as back gate with a thin layer $HfO_2$ on top as insulating gate dielectric layer. The top electrode Ti/Au is directly deposit on top of the monolayer $WS_2$. The thickness of monolayer $WS_2$ is 0.62nm. Our calculation indicates that the absorption and reflectance can be electrically tuned in the range of 40-80% when the charge carrier is tuned from $-0.9 \times 10^{12}$ cm$^{-2}$ to $-13.7 \times 10^{12}$ cm$^{-2}$. The details of the design principle can be found in supplementary information (S2).

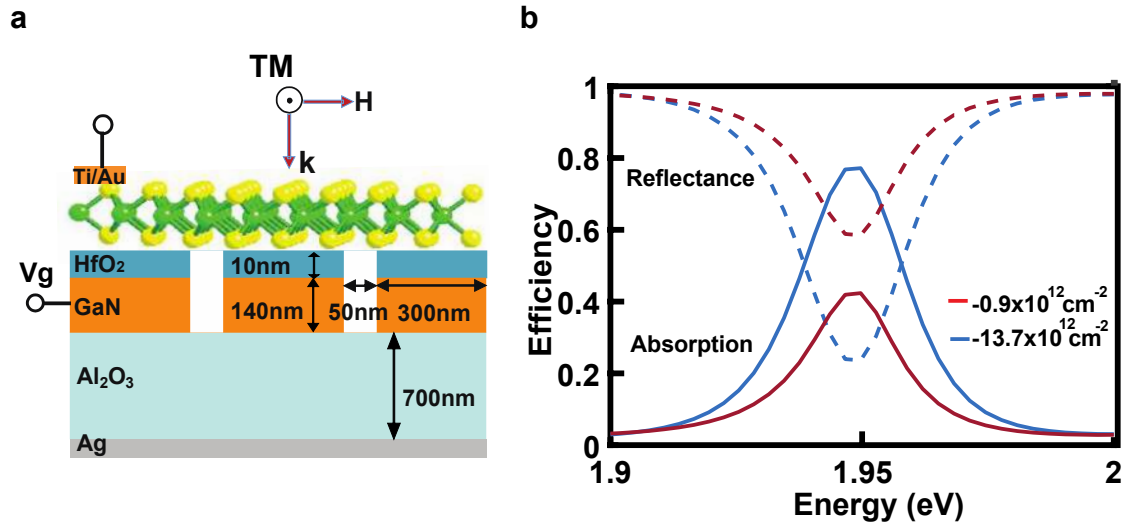

Fig.5 Gating optical functionality. (a) Illustration of configuration of the photonic device. (b) The simulated absorption efficiency (solid line) and reflectance (dash line) correspond to different electron injection induced by gate voltages.

In conclusion, we have demonstrated giant gating tunability in the refractive index (> 60% in the imaginary part and > 20% in the real part) around exciton resonances of atomically thin TMDC monolayers. We believe even larger tunability can be achieved with optimization in the device fabrication. Additionally, we have elucidated that the tunabilitiy mainly results from the spectral broadening ($\gamma$) and the interconversion of the neutral and charged excitons caused by the injected charge carriers. In contrast, the other effects of the injected charge carriers, including bandgap renormalization and change in exciton binding energy, only play negligible roles. The result provides new insight into the fundamental optical properties of 2D TMDC materials. For instance, changing the substrate and dielectric environment of 2D TMDC materials, which is expected to change the screening of Coulomb interactions, may not affect the dielectric function of 2D TMDC materials much[16] unless the changing may induce substantial change in the exciton spectral width or the doping to the materials.[46] More importantly, this result may open up a new age of field-effect photonics whose optical functionality can be electrically controlled in ways similar to that of state-of-art CMOS circuits. We also demonstrate by combining with the nano-photoic structure, the weak absorption nature of 2D TMDC could be overcome and lead to 40% absorption/reflection modulation.

**Acknowledgements**

This work was supported by the National Science Foundation under the grant of ECCS-1508856. The theoretical analysis and discussion are supported as part of the Center for the Computational Design of Functional Layered Materials, an Energy Frontier Research Center funded by the U.S. Department of Energy, Office of Science, Basic Energy Sciences under Grant No. DE-SC0012575. The authors acknowledge the use of the Analytical Instrumentation Facility (AIF) at North Carolina State University, which is supported by the State of North Carolina and the National Science Foundation.

# Giant Gating Tunability of Optical Refractive Index in Transition Metal Dichalcogenide Monolayers

Yiling Yu[1,2], Yifei Yu[1], Lujun Huang[1], Haowei Peng[3], Liwei Xiong[1,4] and Linyou Cao[1,2]*

[1]Department of Materials Science and Engineering, North Carolina State University, Raleigh NC 27695; [2]Department of Physics, North Carolina State University, Raleigh NC 27695; [3]Department of Chemistry, Temple University, Philadelphia PA 19405; [4]Hubei Key Laboratory of Plasma Chemistry and Advanced Materials, Wuhan Institute of Technology, Wuhan, PR China, 430205.

* To whom correspondence should be addressed.

Email: lcao2@ncsu.edu

This PDF document includes

Experimental Methods

Figure S1-S8

S1. Fitting dielectric function using a multi-Lorentzian model

S2. Design optical absorption modulator

Table S1

**Experimental methods**

The Raman and PL measurement were carried out at Horiba Labram HR800 system with incident wavelength of 532 nm. The reflection spectra were collected using a home-built setup that consists of a confocal microscope (Nikon Eclips C1) connected with a monochromator (SpectraPro, Princeton Instruments) and a detector (Pixis, Princeton Instruments). A broadband Halogen lamp was used as incident light for the reflection measurements. The reflectance from the sample is calculated by normalizing the light reflected from sample with respect to the light reflected from a dielectric mirror under the same configuration.

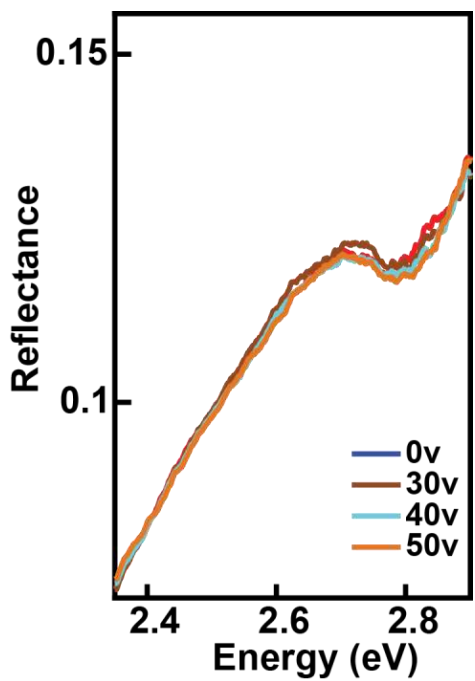

**Figure S1** Negligible dependence in the reflection efficiency of the C exciton of monolayer $WS_2$ on electrical gating.

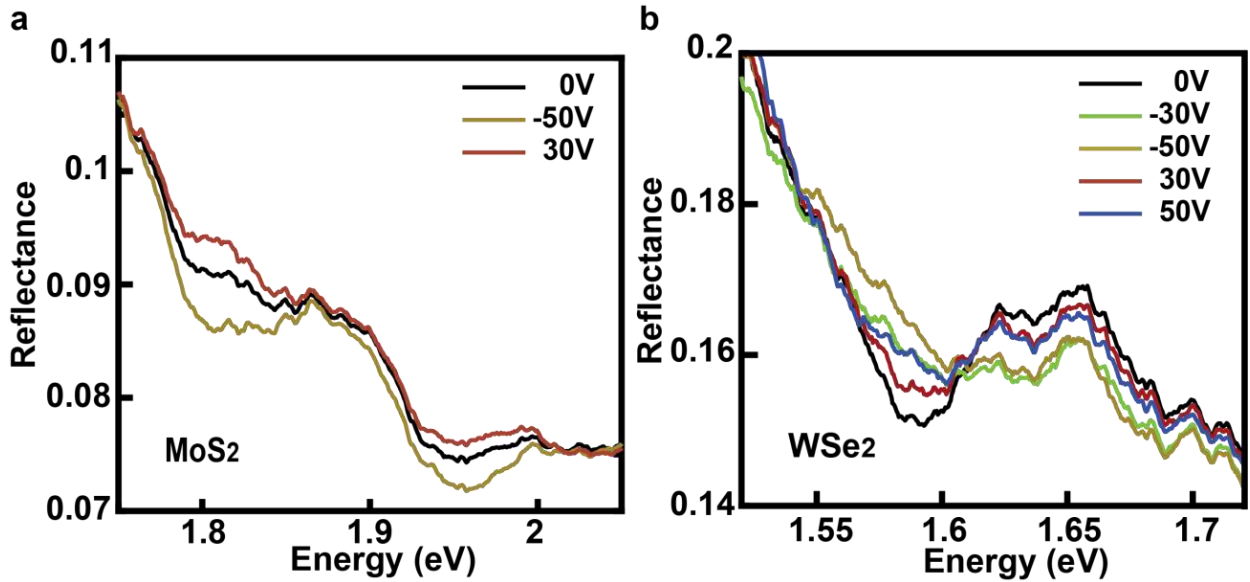

**Figure S2.** Spectral reflection collected from (a) monolayer $MoS_2$ and (b) monolayer $WSe_2$ under different gating voltages.

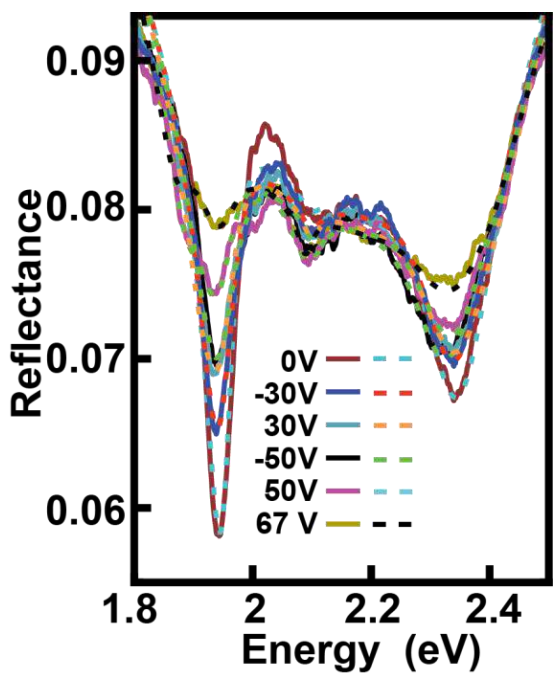

**Figure S3.** Fitted (dash lines) and measured (solid lines) spectra reflection of monolayer $WS_2$ under different gate voltages. In this fitting, the dielectric function is fitted using the multi Lorentzian model.

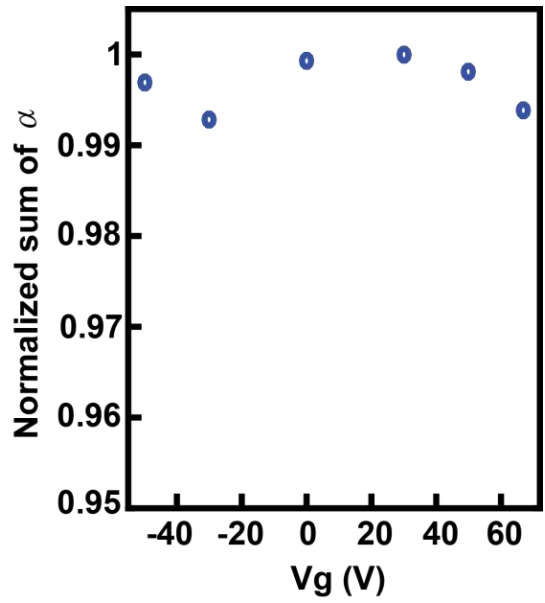

**Figure S4.** The sum of the absorption coefficient $\int_0^{6eV} \alpha(\omega)$ at different gating voltages. The result is normalized with respect to the sum of absorption coefficient at 0V.

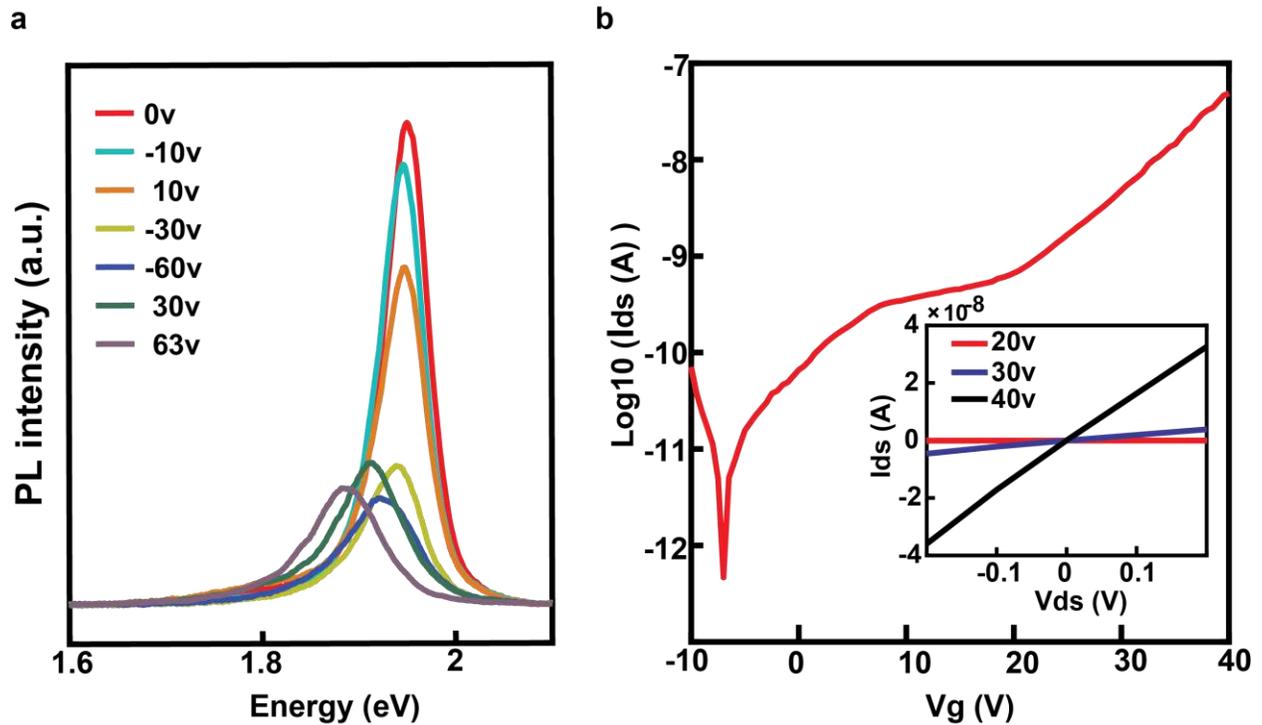

**Figure S5.** (a) Photoluminescence spectra of monolayer $WS_2$ under different gate voltages. (b) I-V curve measurement of monolayer $WS_2$. The inset is I-V curve of drain-source electrodes under different back gate voltages (Vg). The linear relation indicates good Ohmic contacts.

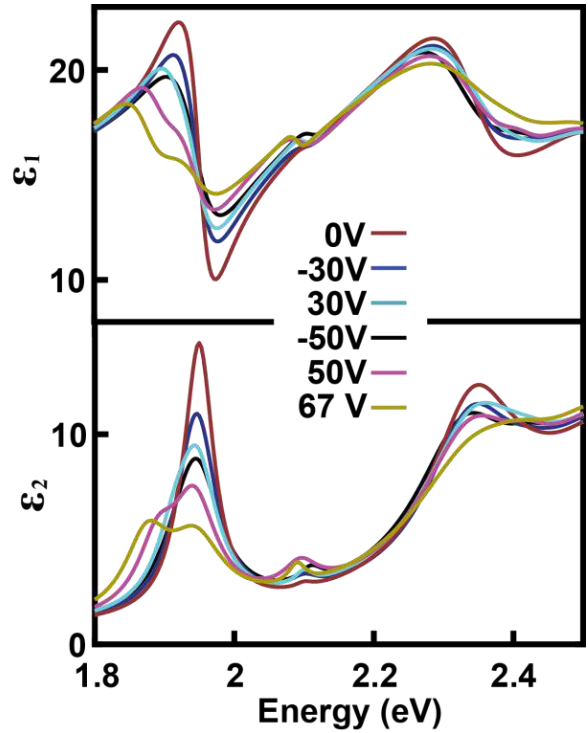

**Figure S6** Fitted real (upper) and imaginary (lower) parts of dielectric function under different gate voltages.

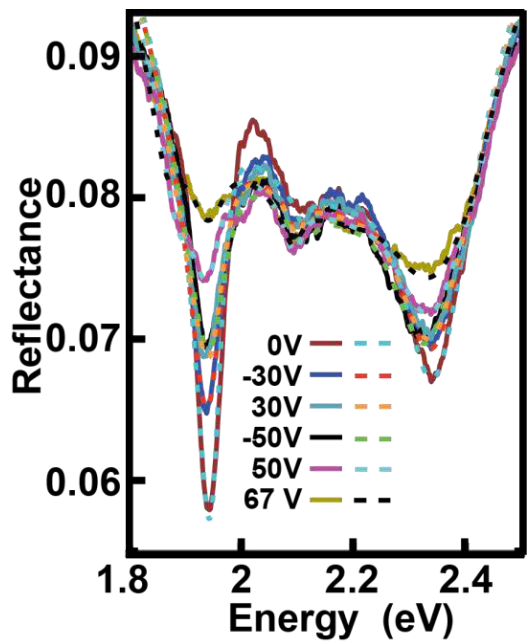

**Figure S7.** Fitted (dash lines) and measured (solid lines) spectra reflection of monolayer $WS_2$ under different gate voltages. The dielectric function is fitted using the Fractional dimensional space model.

## S1. Fitting dielectric function using a multi-Lorentzian model

In order to get accurate determination of the dielectric function $\varepsilon$, the Kramers-Kronig constrained analysis requires information in the full spectral range, but the measured spectral reflection only cover the range of 1.8-2.5 eV. To address this issue, we ignore the contribution from the oscillators in lower energy ranges as it is expected to be weak for the refractive index in the visible range. However, the contribution from the oscillators at higher energy ranges has to be considered. We assume that the dielectric function of the monolayer at the higher energy ranges is similar to that of bulk counterparts, and use the dielectric function of the bulk counterparts, which is available in reference (Ref. 27 in the main text), to correct the oscillators of the monolayers in the higher energy range. Multiple oscillators are set with equal space of 0.1eV and almost equal damping constant 0.3eV. The oscillation strength of these oscillators is fitted to match the dielectric function of bulk $WS_2$ in the UV frequency range. The high frequency Lorentzian oscillators are fitted up to 6eV. The contribution from even higher frequency (larger than 6ev and up to infinite frequency) oscillators are put into $\varepsilon_\infty$. Different sets of oscillation parameters and $\varepsilon_\infty$ have been evaluated to get good matches to both the measured reflection spectrum in visible range (Fig. 1a) and refractive index of bulk $WS_2$ in UV frequency range.

To further exam the accuracy of the fitting method, we measured the refractive index of monolayer $WS_2$ film using standard spectral ellipsometry, and compared it to the refractive index obtained from the fitting of spectral reflection. The two sets of refractive index show nice consistence, with a difference 0f 0.3 and 0.2 in the real and imaginary parts around exciton resonance as indicated by Fig. S8.

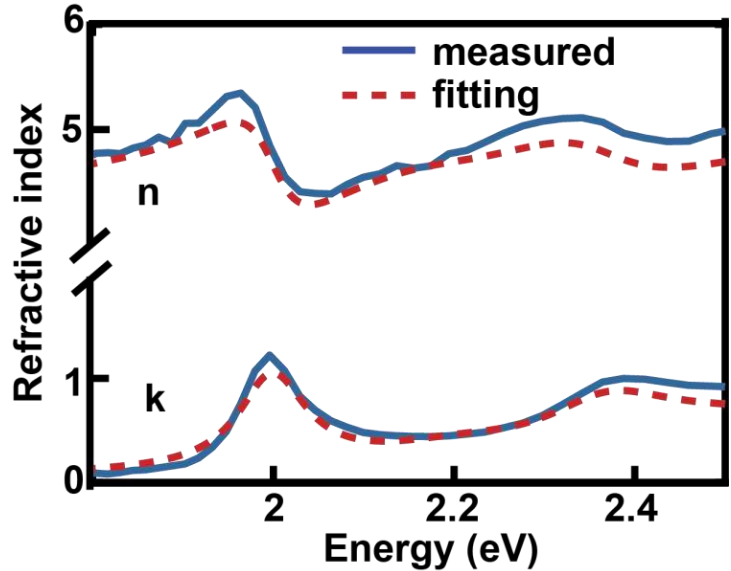

**Figure S8** Measured (blue solid lines) and fitted (red dash lines) refractive index of monolayer $WS_2$ film, (upper) real and (lower) imaginary parts of the refractive index.

Table S1. Parameters used in the calculation for the thermal equilibrium of neutral and charged excitons

| | | | | | | |
|---|---|---|---|---|---|---|
| $n_{+/-}$ (cm$^{-2}$) | $+8.1\times10^{12}$ | $+4.5\times10^{12}$ | $-9\times10^{11}$ | $-6.5\times10^{12}$ | $-10.3\times10^{12}$ | $-13.7\times10^{12}$ |
| $E_{A0}-E_{A+/-}$ (meV) | 23.0 | 13.0 | 24.0 | 37.0 | 53.0 | 65.0 |
| $E_F$ (meV) | 20.8 | 11.6 | 3.0 | 20.7 | 32.8 | 43.6 |
| $E_{b+/-}$ (meV) | 2.2 | 1.4 | 21 | 16.3 | 20.2 | 21.4 |

Note: '+' denote hole doping and '-' denote electron doping. The unit is meV. The binding energy of the charged A exciton can be calculated as $E_{A0} - E_{A+/-} - E_F$, where $E_{A0}$ and $E_{A+/-}$ are the optical bandgap of the neutral and charged A excitons, respectively. $E_F$ is the fermi energy shift with respect to the minimum of conduction band caused by the injected charge carriers. It can be calculated from the charge density $n$ and the density of state in 2D system as $E_F = \hbar^2\pi n/2m^*$. The effective electron mass $0.35m_0$ and effective hole mass $0.46m_0$ are used for calculating $E_F$.[S1] $m_0$ is the free electron mass.

## S2. Design optical absorption modulator

The device design is focused on the spectral range around the A excitonic peak, which is 1.956 eV. Without losing generality, we use two charge densities as examples to illustrate the device design, $-0.9\times10^{12}$ cm$^{-2}$ and $-13.7\times10^{12}$ cm$^{-2}$. The refractive index of monolayer WS$_2$ in the target spectral range (1.956 eV) is 4-1.698i at $-0.9\times10^{12}$ cm$^{-2}$ and 3.86-0.6885i at $-13.7\times10^{12}$ cm$^{-2}$. The design follows a theoretical frame of leaky mode coupling that we have previously developed. [Ref 45]. To ensure the optical resonance in the wavelength of interest, ~ 600nm, we choose 140nm thick GaN nanowire array on sapphire substrate with Ag mirror coated on the back side). 10nm thick HfO2 is deposit on top of GaN as gate dielectrics. The monolayer WS$_2$ may be transferred onto the top of the nanostructure (Fig. 5(a)). The refractive index for the HfO$_2$ and GaN are 2.1 and 2.35, respectively. Our simulation results (Fig. 5(b)) show that the absorption efficiency around 1.956eV is 80% at condition1 (carrier concentration $-0.9\times10^{12}$ cm$^{-2}$) but with only 40% absorption efficiency for condition2 ($-13.7\times10^{12}$ cm$^{-2}$). This really suggests a large absorption and reflection modulation can be achieved by electrical gating monolayer WS$_2$.